\documentclass{sig-alternate}

\usepackage{amsmath}
\usepackage{times}
\usepackage{graphicx}
\usepackage{color}
\usepackage{xspace}

\newcommand{\ignore}[1]{}
\newcommand{\comment}[1]  {}

\newcommand\etal{{\textsl{et al.\,}}}
\def\BE{\begin{equation}}
\def\EE{\end{equation}}
\def\BEA{\begin{eqnarray}}
\def\EEA{\end{eqnarray}}
\newcommand{\cut}[1]{{}}

\newtheorem{thm}{Theorem}

\newcommand{\mdist}[0]{{\em Market Distribution}\xspace}
\begin{document}

\conferenceinfo{SYSTOR 2010}{May 24-26, Haifa, Israel}


\title{A Hybrid Multicast-Unicast Infrastructure for Efficient Publish-Subscribe  in Enterprise Networks}

\numberofauthors{1}
\author{
%
%
\alignauthor
Danny Bickson, Ezra N. Hoch, Nir Naaman, Yoav Tock\\
       \affaddr{IBM Haifa Research Lab,}\\ 
       \affaddr{Mount Carmel, Haifa 31905, Israel}\\
       \email{\{dannybi,ezrah,naaman,tock\}@il.ibm.com}\\
}
\date{10 April 2010}

\maketitle

\begin{abstract}
One of the main challenges in building a large scale
publish subcribe infrastructure in an enterprise network, is to
provide the subscribers with the required information, while
minimizing the consumed host and network resources. Typically,
previous approaches utilize either IP multicast or point-to-point
unicast for efficient dissemination
of the information. 

In this work, we
propose a novel hybrid framework, which is a combination of both
multicast and unicast data dissemination. Our hybrid framework
allows us to take the advantages of both multicast and unicast,
while avoiding their drawbacks. We investigate several algorithms
for computing the best mapping of publishers' transmissions into
multicast and unicast transport.

Using extensive simulations, we
show that our hybrid framework reduces consumed host and network resources, outperforming traditional solutions. To insure the subscribers interests closely resemble those of real-world settings, our simulations are based on stock market data and on recorded IBM WebShpere subscriptions.\end{abstract}


Performance


\section{Introduction}

Consider a large-scale publish-subscribe application that is
characterized by a large number of information flows, as well as a
large number of subscribers. Each information flow generates
messages which must be delivered to an interested subset of
subscribers. Subscribers are interested in different, yet possibly
overlapping, subsets of the information flows. Naturally, an
individual information flow may be required by many subscribers.
A typical example is a financial market data dissemination
system, where the flows can be stock quotes (of which there are tens of
thousands), commodity prices, etc., and subscribers are
traders, analysts and so on (in hundreds). Each subscriber is
interested in a different portfolio.

One of the two common approaches in the above dissemination
scenario is to utilize IP multicast to transmit the data.
In this work we assume that IP multicast service is supported in the enterprise network.
Given that overlaps between subscribers'
interests are not rare, transmission costs can be reduced by
grouping information flows into groups, and using multicast to
disseminate these flows to subscribers. This mechanism requires
two mappings: one between flows and groups (mapping of a flow to one or more
multicast group), and another mapping between users and multicast
groups (such that each subscriber gets all the information she is interested in).
The problem of finding the mappings which minimize consumption of network
resources is termed {\em ``the channelization problem''}
\cite{Adler-01}.

Using multicast as a mean of dissemination has some limitations.  Typically, there is a
limited amount of multicast addresses which can be used, due to the memory and
computational overhead of the network devices.

In our setting, the number of flows is much larger than the number of available
multicast groups, which means that a one-to-one mapping of flows
to multicast groups is not possible. Thus, different flows have to
be batched into the same multicast group. As a result subscribers
may receive data they are not interested in and which
they must filter. As shown in~\cite{Carmeli-04}, filtering in the end hosts is one
of the reasons for reduced performance in a high bandwidth enterprise network.

A second common approach is to use point-to-point connections,
where each publisher transmits the information required using unicast. The main drawback of solely using
unicast is the poor utilization of network resources when
many subscribers are interested in the same data flow. In this
case, the transmitter has to transmit the same data many times to
different users, which results in a waste of transmitter resources like bandwidth, CPU and memory
as well as wasted network bandwidth.

In the current paper, we propose a novel hybrid approach, which combines both
unicast and multicast transports. In our approach, we allow a flexible allocation
of flows into unicast and multicast connections, avoiding the inherent drawbacks
of using a single scheme. Thus, we gain high efficiency when many subscribers are
interested in the same data flow by utilizing multicast, and use unicast to reduce
unneeded filtering, whenever the subscription to certain flows is relatively rare.

We define the hybrid unicast-multicast problem as an optimization problem,
and explore several heuristics to solve it. Using extensive simulations, we compare different
approaches for solving the hybrid problem and identify which perform best, under different
scenarios. As an additional contribution, we explore different algorithms for solving
the related channelization problem, which is NP-hard, and identify a single algorithm which outperforms the others.

The paper is organized as follows. Section~\ref{sec:rel_work}
overviews the related work and explains the novelty in our hybrid approach.
Section~\ref{sec:model} describes the problem model and formally defines
the hybrid problem, showing it is a NP-Hard problem. Section~\ref{sec:algos} presents our proposed heuristics for solving the hybrid problem. Section~\ref{sec:exp_res} gives extensive
experimental results which compare the different heuristics and their operation under
various real-world scenarios. We conclude in Section~\ref{sec:conc}.

\section{Related Work} \label{sec:rel_work} Publish-subscribe systems have been the
target of extensive research in the last years. Research has focused on
the problem of disseminating data efficiently to interested users.
Two main paradigms were proposed: content-based multicast and subject-based multicast~\cite{Levine-00,citeulike:3385825,Lety-04}). Different extensions to the paradigms
include~\cite{10.1109/ICDCS.2005.42} where a hybrid approach for
content-based and subject based dissemination is proposed. Another
example is~\cite{1069879} which proposes a solution for a setting
in which dynamic changes of the multicast groups is required. In~\cite{Opyrchal-00} content-based dissemination is implemented using IP multicast.

%

One of the main challenges when considering subject-based
multicast is in solving the channelization problem
(\cite{Adler-01, Wong-99, citeulike:3385726, IBMkmeans}). Previous
approaches map flows into multicast groups, while the current
paper allows for both multicast and unicast transmissions. In
Section~\ref{sec:exp_res} we empirically compare several
algorithms for solving the channelization problem, identifying a
single algorithm which outperforms the others.

A closely related work to ours is Dr. Multicast \cite{DrMulticast}
which proposes to use unicast as well as multicast in a data
center information dissemination scenario. However,
\cite{DrMulticast} focuses on the management and stability issues
of IP multicast in the data-center, and does not explicitly
explore the question of mapping flows into multicast and unicast
in a quantitative manner. To the best of our knowledge, we are the
first work which formally defines the problem as an optimization
problem, and explores several heuristics to solve it.

\section{Model and Problem Definition} 
\label{sec:model}
We use the following notations as in~\cite{Adler-01}.

\begin{itemize}
  \item Let $m$ be the number of users.

  \item Let $n$ be the number of flows.

  \item Let $k$ be the number of mulitcast groups.

  \item Let $W_{n \times m}$ denote the interest matrix:
      \begin{eqnarray*}
      W_{i,j} &=& \left\{
            \begin{array}{ll}
              1 & \textrm{user } j \textrm{ is interested in flow  } i\\
              0 & \textrm{otherwise} \\
            \end{array} \right.
      \end{eqnarray*}

  \item Let $X_{n \times k}$ denote the mapping from flows to multicast groups:
      \begin{eqnarray*}
      X_{i,j} &=& \left\{
            \begin{array}{ll}
              1 & \textrm{flow } i \textrm{ is mapped to multicast group  } j \\
              0 & \textrm{otherwise} \\
            \end{array} \right.
      \end{eqnarray*}

  \item Let $Y_{k \times m}$ denote the mapping from multicast groups to users:
      \begin{eqnarray*}
      Y_{i,j} &=& \left\{
            \begin{array}{ll}
              1 & \textrm{user } j \textrm{ receives multicast group  } i \\
              0 & \textrm{otherwise} \\
            \end{array} \right.
      \end{eqnarray*}

  \item Let $T_{n \times m}$ denote the unicast matrix:
    \begin{eqnarray*}
      T_{i,j} &=& \left\{
            \begin{array}{ll}
              1 & \textrm{flow } i \textrm{ is sent to user  } j \textrm{ using unicast} \\
              0 & \textrm{otherwise} \\
            \end{array} \right.
      \end{eqnarray*}
  \item Let $\lambda_{1 \times n}$ denote the rate of the flows where $\lambda_{i}$ is the rate of flow $i$.

\end{itemize}


%

\subsection{The Channelization Problem}
\begin{figure}[ht!]
\begin{center}
  \includegraphics[scale=0.7]{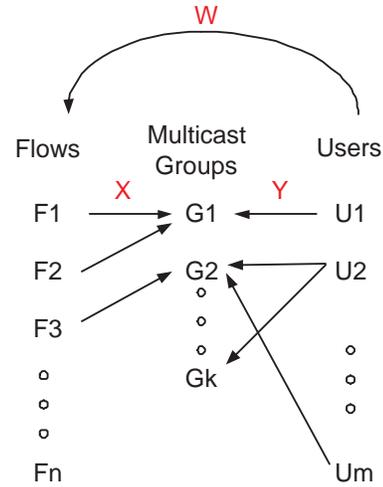}\\
  \caption{Schematic picture of the channelization mapping.}\label{fig:channelization}
\end{center}
\end{figure}

Given $m$ users, $n$ flows, $k$ multicast
groups, a vector of flow rates $\lambda$ and an
interest matrix $W$, the channelization problem~\cite{Adler-01} aims at finding two mapping matrices $X, Y$ that minimize the cost of transmission (using only multicast groups), under the constraint that each user receives all the flows it is interested in. A schematic diagram of the channelization mappings is given in Figure~\ref{fig:channelization}.

To formally define the cost function, let $w_1, w_2$ be real positive numbers,
\begin{eqnarray*}
  C(X, Y) & = & w_1 \sum_{i=1}^{n}\sum_{j=1}^k \sum_{h=1}^m X_{i,j}Y_{h,j}\lambda_i + \\
  && w_2 \sum_{i=1}^n \sum_{j=1}^k X_{i,j} \lambda_i\;.
\end{eqnarray*}

The cost consists of two terms; the first sums all transmission received by subscribers. For each user $h$ it sums the number of times $h$ receives any given flow $i$, times the rate $\lambda_i$ of flow $i$. The second term sums the transmission of the senders; that is, each flow $i$ is summed according to the number of multicast groups it is transmitted to, times the rate $\lambda_i$. $w_1, w_2$ are factors which weight the two terms.
The channelization problem is defined as:
\[ \min_{X,Y} C(X) \]
\[ \mbox{s.t.}\ \  XY \ge W\,. \]
In other words, given a set of users $U$, a set of multicast groups $M$, a set of flows $F$, an interest matrix $W$ and a flow-rate vector $\lambda$; find $X, Y$ that minimize $C(X, Y)$ under the constraint that $XY \geq W$.

\subsection{The Hybrid Channelization Problem}
Below we model our hybrid framework as an optimization problem. Unlike the original channelization problem, the transmitters may send flows using unicast. That is, any flow $f$ can be transmitted using unicast to any user $h$.
In the hybrid problem the cost function $C'$ is composed of three terms:
\begin{eqnarray}
  C'(X, Y, T) & = & w_1' \sum_{i=1}^{n}\sum_{j=1}^k \sum_{h=1}^m X_{i,j}Y_{h,j}\lambda_i + \nonumber \\
  && w_2' \sum_{i=1}^n \sum_{j=1}^k X_{i,j} \lambda_i + \nonumber \\
  && w_3'(w_1'+w_2') \sum_{i=1}^n \sum_{h=1}^m T_{i,h} \lambda_i
  \label{HC}
\end{eqnarray}

The additional term represents the cost of all the flow $i$-user $h$ pairs such
that flow $i$ is sent using unicast to user $h$, multiplied by the cost of
transmission. The cost of transmission of a flow consists of the
cost of sending the flow ($w_2'$), the cost of receiving the flow
($w_1'$), and the cost incurred by the unicast mechanism, $w_3'$
(additional memory requirements, etc). For the rest of the paper,
we assume that the transmitting and receiving costs are equal
($w_1'=w_2'$) and that the unicast cost equals their sum ({\it i.e.}, $w_3' =1$).

Using the cost function $C'(X, Y, T)$ the hybrid channelization
problem can be formally defined. Given $m$ users, $k$ multicast
groups, $n$ flows, an interest matrix $W$ and a flow-rate vector
$\lambda$, the hybrid channelization problem is defined as: 

\[ \min_{X,Y,T} C'(X, Y, T) \]
\[ \mbox{s.t \ \ } XY + T \geq W\,.\] The constraint $XY + T \geq W$ requires that each user $h$ requesting flow $i$ will
either receive $i$ via unicast or via a multicast group $h$
listens to. ($h$ may receive flow $i$ via both multicast and unicast; however, that would be wasteful.)

\subsection{Intractability of the
Hybrid Channelization Problem}\label{sec:nphard}
\begin{thm}
The hybrid channelization problem is NP-Hard.
\end{thm}
\begin{proof}
 In \cite{Adler-01} it was shown that the non-unicast problem is NP-Hard,
therefore the unicast channelization problem can be reduced to the
non-unicast channelization problem as a proof of its hardness. The
reduction is simple: given $n,m,k,W, w_1, w_2,\lambda$ as
input to the channelization problem, construct an
input to the hybrid channelization problem which is exactly the
same, with a single modification. Setting $w_3'$ to be large than
$C(1_{n \times k}, 1_{k \times m})$ ensures that any
solution $X, Y, T$ does not have a lower cost than $X, Y, 0_{n
\times m}$. Thus, the minimal cost is the same as in the
non-unicast setting. 
\end{proof}

\section{Proposed Algorithms} \label{sec:algos}
We propose the
following two-step framework for solving the hybrid problem. In
the first step, we solve the channelization problem,
without assigning any unicast flows. In the second step, we
heuristically select some of the flows to be sent using
unicast, and update the subscription matrix $W$ accordingly.

This process can be repeated several times, as long as the system
cost is reduced. Simulation results show that repeating the
process does not
significantly improve system cost, while having a high
computational cost.

The above process can be viewed as starting from a solution that
uses only multicast, and then using unicast to greedily improve
the solution. Alternatively, one can start with a solution that
uses only unicast ({\it i.e.}, $T=W$), and greedily improve it by
moving flows to multicast. Our simulations show that the first
framework operates better than the latter one; while both of
them improve upon previous non-hybrid solutions.


\subsection{First Step: Solving the Channelization \\ Problem}
Previous work ~\cite{IBMkmeans,Adler-01} discuss several heuristics for
solving the channelization problem. Adler \etal examine several
heuristics, among them, random assignment and user and flow based
merges. Tock \etal proposed a variant of the K-Means algorithm
which greedily minimizes the cost on each iteration.

In this work, we extensively compare the different previous approaches, while exploring new algorithms.
We have utilized an algorithm from the data mining domain, called Binary Matrix Decomposition (BMD \cite{BMD,BMD0}) which is originally used for two-sided binary clustering of documents and keywords into document classes.
The basic idea is that the global cost function for minimization is:
\[ \min_{X,Y} ||XY - W||^2_2 \]
\[ \mbox{subject to  } X,Y \in \{0,1\} \]
which means we are looking for a decomposition of the
interest matrix $W$ into two binary matrices $X,Y$ so that the
Euclidian norm between $XY$ and $W$ is minimized. An alternating
algorithm is derived by starting with an initial guess $X$,
solving $Y$ which minimizes the cost function and then continuing
in rounds. There are some drawbacks in using this algorithm:
first, it does not support variable flow rates. Second, it
allows for some flows to be missing. Despite those drawbacks it has
reasonable performance when operating on small systems ({\it i.e.}, 200 flows, 10 multicast groups, 100 users).
However, when operating on larger systems ({\it i.e.}, 10000 flows, 100 multicast groups, 250 users) it does not improve upon a random selection of a solution. Therefore, we have omitted the experimental results of
the BMD algorithm from the following graphs.

We have also utilized the Matlab\texttrademark \ K-Means algorithm \cite{matlabkmeans1,matlabkmeans2} which is
a two phase algorithm. In the first phase points are
reassigned to their nearest cluster centroid, all at once,
followed by recalculation of cluster centroids. The second phase
uses ``on-line'' updates, where points are individually reassigned
while reducing the total cost function, and cluster centroids are
recomputed after each reassignment.

We further investigated an interior point algorithm. Starting from
the original problem formulation by Adler \etal, the binary mapping
matrices $X$ and $Y$ are relaxed to be in the range $(0,1)$. The
constraints that $X > 0, Y > 0, X < 1, Y < 1$ and $XY \geq W$ are
incorporated into the cost function using the log-barrier
technique (\cite{BV04}) and then the Newton method is applied.
After convergence, the solution is rounded to receive binary $X$
and $Y$. Unfortunately, the interior point method did not perform
well in practice. Some of the reasons are that the problem is neither
concave nor convex. We have usually received a good fractional
solution, but when the solution was rounded to the closest
integer solution, it did not compare favorably to the other
algorithms. Therefore, we have omitted the experimental results of
the interior-point algorithm from the following graphs.

In total, we have examined five different algorithms for solving
the channelization problem. Table~\ref{tab:algos} summarizes the
tested algorithms. Regarding their running time, not surprisingly,
the random assignment is the lightest algorithm with running time
of $n$ (setting each flow to a random multicast group) plus $mnk$
for going over all users and assigning them to groups such that
they will receive all required flows. The two K-means variants as
well as the BMD algorithms have a similar running time, where $t$
is the number of iterations (typically five on problem
sizes of thousands), since for each flow they go over all possible
assignments of groups by taking the minimal cost. The interior
point method running time is dominated by the Hessian inversion in the Newton step.


\begin{table}
\begin{center}
\begin{tabular}{|l|l|}
  \hline
  Algorithm & Running time \\ \hline
  Random assignment & $O(n + mnk)$ \\
  K-means \cite{IBMkmeans} & $O(tmnk)$ \\
  Matlab K-means \cite{matlabkmeans1,matlabkmeans2} & $O(tmnk)$ \\
  BMD \cite{BMD,BMD0} & $O(tmnk)$ \\
  Interior-point method & $O(t(n^3 + m^3))$ \\
  \hline
\end{tabular}
\caption{Examined algorithms for solving the channelization
problem and their running time.}
 \label{tab:algos}
\end{center}
\end{table}

\subsection{Second Step: Choosing Flows for Unicast}\label{sec:chooseSys}

Different ways of choosing flow-user pairs can be used. We concentrated on two
different types of heuristics: flow based and user based.
Flow based heuristic means that each flow $i$ is
either sent to all users that are interested in it via unicast, or
transmitted to all of them via multicast; one can either remove the
``heaviest'' flow or the ``lightest'' flow (in the sense of the amount of bandwidth required to transmit that flow to all users interested in it). Clearly, we expect the lightest-flow approach to outperform the heaviest-flow approach; this has been validated by our simulations, and in the following graphs we will consider only the lightest-flow approach.

User based heuristics means that all
flows sent to user $h$ are sent via unicast. That is, if user $h$
receives any flow $i$ using unicast, then any other flow $i'$ that
is sent to $h$ is sent using unicast. Similar to the case of flow
removal, we can choose to remove the ``heaviest'' or ``lightest''
user (here ``heavy'' and ``lightweight'' means the total bandwidth required to transmit all flows user $h$ is interested in). Our simulations show the heaviest-user approach outperforms the lightest-user approach; the reason lies in the fact that heavy users listen to many multicast groups, and thus incur large overhead in filtering. In the following graphs we show the heaviest-user approach only.

To sum up, we have tested the heuristics of removing the
heaviest/lightest flow/user from $W$, and moving it to $T$. In
addition, each of the above options was tested twice: once by
finding a single $X, Y$ pair then removing as many flows/users
from $W$ as possible (termed ``non-iterative''); and once by
finding a $X, Y$ pair, removing a single flow/user from $W$,
then finding a new $X, Y$ pair (that optimizes the modified $W$),
removing another flow/user from the altered $W$, and so on
(as long as the cost function was minimized); termed
``iterative''. Our simulations have shown the non-iterative approach operates almost as good as the iterative, with significantly reduced computational effort. Thus, the following graphs depict only the non-iterative runs.

In addition, we have tested several other heuristics. The basic
idea is to remove a flow/user in a greedy way, {\it i.e.},
repeatedly move to unicast the user/flow/flow-user pair that
minimizes the total cost (Eq.~\ref{HC}), until cost does not
decrease or bandwidth for unicast is fully
utilized\footnote{Without loss of generality, we assume there is a
limit on the total amount of bandwidth allocated for unicast. This
limit is used as a stopping criteria for our algorithm}. We call
those heuristics greedy user, greedy flow and greedy flow-user
pair accordingly.

In practice, the flow-user pair heuristic did not perform well, while having a high computational cost. Thus, it is not shown in the graphs. To sum up, we have tested in total eleven different heuristics. In the following section, we present the simulations' results for the best-performing among these heuristics.

\section{Experimental Results} \label{sec:exp_res}
We have experimented with three possible initializations of the user
interest matrix $W$. The first one is {\em Random}, where each user uniformly selects \%3 of the flows. The second one, \mdist, is based on a model of subscription patterns in financial messaging systems~\cite{IBMkmeans}. This model is based on stock market symbol rates collected from the New York Stock Exchange (NYSE). The matrix $W$ was
composed of 10,000 symbols divided into 10 markets, and 250 users. Each user was interested in 4 markets, and chose some of the symbols in each selected market. The flows within a market are distributed exponentially, and the markets are distributed using Zipf distribution. The \mdist determines the flow rate $\lambda$ as well.

%

\begin{figure}[tbp!]
\begin{center}
    \includegraphics[width=0.4\textwidth]{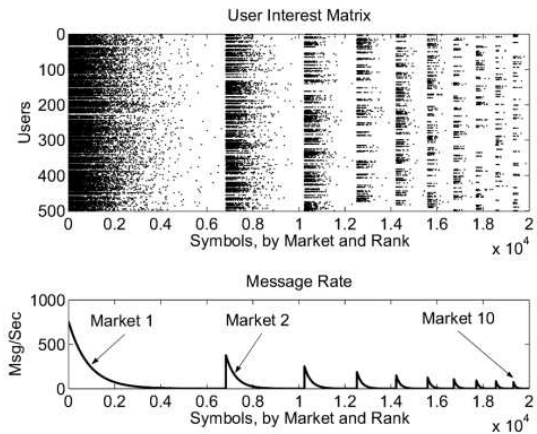}
    \caption{Initialization of user interest matrix $W$ and message rate ~$\lambda$ according to the Market Distribution model.}
    \label{fig:user-interest-matrix}
\end{center}
\end{figure}

Figure~\ref{fig:user-interest-matrix}
shows an example of a user interest matrix (top), and the relative
message rate of each symbol (bottom), according to the \mdist.

The third initialization to the matrix $W$ uses a subscription pattern captured from
an IBM's WebSphere \cite{websphere} test cluster. In it there are 79 processes subscribed to over 6100 topics.
Subscription to the topics is entirely automatic, influenced by the configuration and load incurred upon the cell.

\begin{figure}[tbp!]
\begin{center}
    \includegraphics[width=0.4\textwidth]{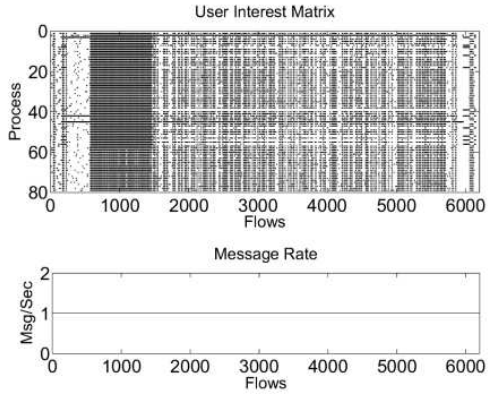}
    \caption{User interest matrix of an IBM WebSphere cell, with automatic subscriptions to topics.}
    \label{fig:user-interest-matrix2}
\end{center}
\end{figure}

As can be seen in Figure~\ref{fig:user-interest-matrix2}, the resulting interest matrix is clearly different from the one generated by the model of human behavior in financial markets (see Figure~\ref{fig:user-interest-matrix}). Importantly, many topics have identical audiences, which perfectly lends itself to multicast channelization.

\subsection{Performance of the different algorithms}

\begin{figure}[h!]
    \includegraphics[width=0.45\textwidth]{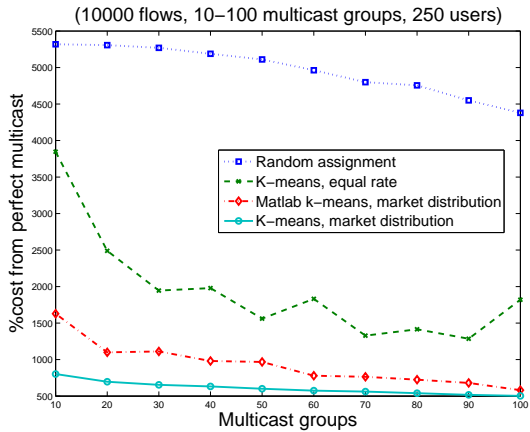}
    \caption{The non-hybrid channelization problem. Both K-means algorithms perform superiorly when rate is equal.}
    \label{fig:non-unicast}
\end{figure}

Among the algorithms listed in Table~\ref{tab:algos}, only
the K-Means and the interior-point method take the flow rates
$\lambda$ into consideration. Thus, only the K-means was plotted twice, once with equal rate and once with rate derived
 by the \mdist, as shown in Figure~\ref{fig:non-unicast}. Using equal rate, both
K-means and Matlab K-means have a superior performance. However, using
\mdist rate, the K-Means algorithm has a noticeably superior performance over all others.

In all graphs shown, the Y axis represents percentage cost from
perfect multicast, where the term perfect multicast refers to the
cost of transmission using multicast transport only (with no
unicast), assuming there are unlimited number of multicast groups.
Thus, perfect multicast means that each user receives exactly all
flows it is interested in, each flow is transmitted only once and
there is zero filtering in the network.

In the hybrid setting, we allow some of the traffic to be
transmitted using point-to-point connections. We have tested
different heuristics for moving traffic from multicast to unicast
(see Subsection~\ref{sec:chooseSys}).

\begin{figure}[h!]
    \includegraphics[width=0.45\textwidth]{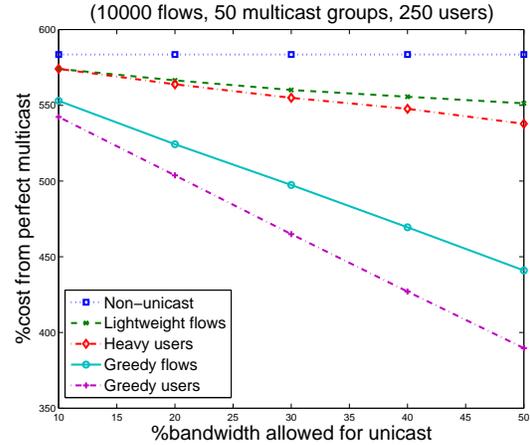}
    \caption{Comparing different heuristics, with clear advantage for the greedy-based algorithms.}
    \label{fig:unicast_methods}
\end{figure}

Figure~\ref{fig:unicast_methods} compares the top heuristics:
lightest-flow, heaviest-user, greedy flow and greedy user. As can
be seen, allowing some of the data to be unicasted reduces the
cost.
Evaluated using the Market Distribution, it seems that the
greedy-user heuristic outperforms the greedy-flow heuristic.
However, this result is overturned when evaluating using the
WebSphere distribution (in the sequel). Thus, the relative
competitiveness of these two heuristics depends of the nature of
the interest matrix.


\subsection{Effect of the Interest Matrix $W$ on performance}
\begin{figure}[h!]
    \includegraphics[width=0.45\textwidth]{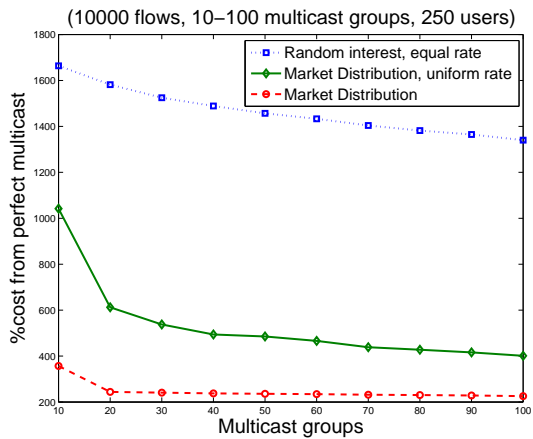}
    \caption{Different interest matrices and their effect on performance.}
    \label{fig:interest_matrix}
\end{figure}

The interest matrix $W$ represents the flows each user is
interested in. The performance of the different heuristics is
highly dependent on the content of $W$ which represents the characteristics of the instance of the problem. In
Figure~\ref{fig:interest_matrix} the lightest-flow heuristic is evaluated with different interest matrices: a random interest matrix where each flow has the same
rate, a \mdist interest matrix where all flow have a fixed same rate,
and a \mdist interest matrix where the rates are also according to \mdist. As can
be seen in Figure~\ref{fig:interest_matrix}, the algorithm
performs best when running on a \mdist interest matrix; {\it i.e.}
the heuristic is optimized for the expected distribution of a real-world financial market application.
This happens because of the underlying Zipf probability, where the top flows are requested
by a large number of users. This makes the clustering of top flows into multicast groups easier.

\begin{figure}[h!]
    \includegraphics[width=0.45\textwidth]{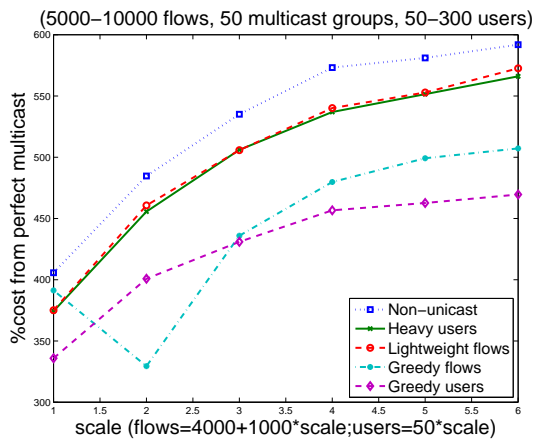}
    \caption{Effect of scaling on performance.}
    \label{fig:scalability}
\end{figure}

Figure~\ref{fig:scalability} shows how the different heuristics perform as the size of the system increases.
Each point in the figure represents a different system: for point $i \in \{1,..., 6\}$, the system consists of $4000 + 1000 \cdot i$ flows, $50 \cdot i$ users while the number of multicast groups is fixed to 50. We did not scale
the number of multicast groups since it is usually dictated by the networking hardware.

The relation between the different heuristics is mostly preserved at different system sizes. An interesting exception is point $i=2$, in which the greedy flows outperforms the greedy users heuristic. This effect is not surprising as different systems (specifically, the ratio between flows, users and multicast groups) can change the relative efficiency of the different heuristics.

\begin{figure}[h!]
\begin{center}
    \includegraphics[width=275pt]{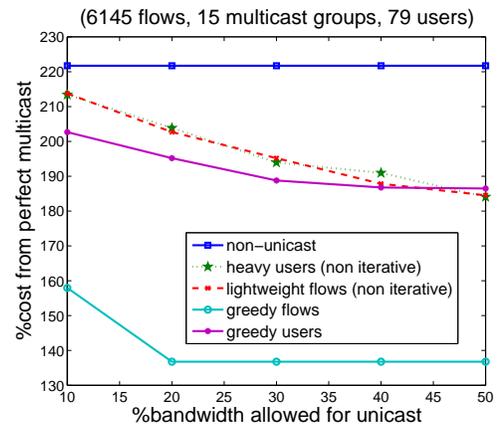}
    \caption{Comparing different heuristics on a trace of a WebSphere cell.}
    \label{fig:unicast_methods2}
\end{center}
\end{figure}

To show the behaviorial difference of the heuristics when running
on a mechanical subscription trace, we have ran the different
heuristics on the IBM WebSphere distribution (see
Figure~\ref{fig:unicast_methods2}). As can be seen, when the
subscription patters closely overlap, the flow based heuristic
outperform the user-based heuristics. It is interesting to note
that the heavy-user heuristic actually increases the cost, since
this heuristic moves the heaviest user and does not consider the
cost of the move. In addition, the greedy user and greedy flow
heuristics reach their maximal improvement at very low unicast
bandwidth. This phenomena is due to the structured nature of the
interest matrix, incurred by the mechanical subscription pattern.

\begin{figure}[h!]
    \includegraphics[width=275pt]{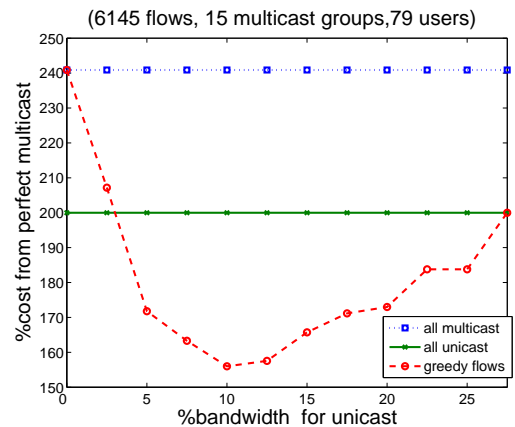}
    \caption{System cost for the hybrid approach using the greedy flow heuristic. }
    \label{fig:unicsat0-100}
\end{figure}

Figure~\ref{fig:unicsat0-100} represents well the benefits of using our hybrid approach.
The greedy heuristics is forced to use a given percentage of unicast
bandwidth (the X-axis), using the WebSphere subscription pattern. Using the hybrid approach, the greedy flow heuristic improves upon both the multicast only and unicast only schemes. The total cost of transmission is reduced in a way which is not possible using a single scheme.

\subsection{Discussion}
We have experimented with different heuristics for selecting
which of the data should be transmitted using unicast. Under the
stock market model, the best heuristics are greedy heuristics
which repeatedly move a single user/flow from multicast to unicast
to minimize the total cost. In this setting, the distribution
leads to several multicast groups which carry a large number of
non-identical heavy flows. Thus, a user that is interested in any
heavy flow might be forced to receive it via a multicast group
that carries other heavy flows that he does not need, leading to a
high filtering cost.
In this case, the gain of moving a single user to unicast
outweighs the gain that might be achieved by moving the best flow
to unicast, because the best flow to be moved is usually fairly
light weight. Therefore, the heuristic of greedily moving users
from multicast to unicast works well in this setting.

The second scenario we tested consisted of a user interest matrix
from a WebSphere test cluster. As the users of this problem are
software / script based, their interests are homogenous. Thus,
many users can use the same multicast group with no need for
filtering. Therefore, a flow which is of interest to a few users
can incur heavy cost on filtering, if it is  assigned to a
multicast group that many users listen to. This property causes
the user based heuristics to perform poorly, while the flow based
heuristics perform well.

In other words, the greedy-user and greedy-flow schemes
``thin-out'' the interest matrix by removing rows and columns,
respectively, making the resulting interest matrix more amenable
to channelization. The relative competitiveness of these
heuristics depend on the structure of the interest matrix.

\section{Conclusion} \label{sec:conc} This paper analyzes the
hybrid channelization problem. We formally define the problem as
an optimization problem and propose efficient heuristics for
solving it. Our general framework starts from a solution to a
non-unicast problem and tries to improve it by allowing some of
the data to be transmitted via unicast. Similarly, we start from a
solution which utilizes only unicast, then improve it by allowing
some of the data to be transmitted via multicast.

We have tested our heuristics against two different real-world
scenarios. First is a simulated brokers' interest in financial
market data and the second is mechanical subscription pattern
captured from an IBM WebSphere test cluster. Five different
algorithms for solving the non-unicast channelization problem
where examined, and a single algorithm, the K-means algorithm was
identified to perform the best in all settings.

In total we have experimented with eleven different heuristics.
The greedy heuristics (that improve the cost function directly) performed
better than the others. However, greedy heuristics should be taken with a salt of grain,
as different problems incur different distributions on the user
interest matrix $W$ and on the rate of the flows. Thus, different
heuristics may perform differently as the problem context changes.

To conclude, by allowing a combination of multicast and unicast
transmissions, we gain in reduced host and network resource
consumption. It seems that the performance of a publish subscribe
system is highly depended on the subscription patterns. Our
hypothesis is that user based heuristics combined with the flow
based heuristics cover a large range of problems. Thus, we provide
a range of heuristics that can be used to practically deploy a
publish subscribe system efficiently.









\bibliographystyle{abbrv}

%
%

\end{document}